\documentclass[11pt]{article}
\usepackage{moriond,epsfig}

\newcommand{\eqnlessref}[1]{(\ref{#1})}

\def\scrD{{\cal D}}
\def\lsim{\mathrel{\lower0.3em\hbox{$\stackrel{\textstyle <}{\sim}$}}}
\def\gsim{\mathrel{\lower0.3em\hbox{$\stackrel{\textstyle >}{\sim}$}}}

\def\negspace{\kern -0.4em}

\def\dvec{\raise 0.3 em \hbox{$^\leftrightarrow$} \kern -0.77 em}

\def\omegahat{\hat%
	{\setbox0=\hbox{$\omega$}%
		\kern-.025em\copy0\kern-\wd0
		\kern.05em\copy0\kern-\wd0
		\kern-.025em\raise.0433em\box0}}

\begin{document}
\vspace*{4cm}
\title{Effective String Theory of Dual Superconductivity}

\author{M. Baker and R.Steinke}
\address{Department of Physics, University of Washington, P.O. Box 351560, Seattle, WA 98195, USA\\}

\maketitle\abstracts{
    We show how an effective field theory of long distance QCD,
    describing a dual superconductor,
    can be expressed as an effective string theory of superconducting
    vortices. We use the semiclassical expansion of this effective
    string theory about a classical rotating string solution
    in any spacetime dimension $D$ to obtain the semiclassical meson 
    energy spectrum.
    For $D=26$, it
    formally coincides with the energy spectrum of the open string in classical
    Bosonic string theory. However, its physical origin is different.
    It is a semiclassical spectrum of an effective string theory valid
    only for large values of the angular momentum.  For $D=4$, 
    the first semiclassical correction adds the constant $1/12$
    to the classical Regge formula for the angular momentum of
    mesons on the leading Regge trajectory.} 


\section{The Dual Superconductor Mechanism of Confinement}

In the dual superconductor mechanism of 
confinement~\cite{Nambu,Mandelstam,tHooft} a dual Meissner
effect confines color electric flux 
to narrow tubes~\cite{Nielsen+Olesen} connecting a quark-antiquark pair. 
In the confined
phase monopole fields $\phi$ condense to a
value $\phi_0$, and dual potentials $C_\mu$
acquire a mass $M = g\phi_0$ via a dual Higgs mechanism.
The dual coupling constant is $g=2\pi/e$, where $e$ is
the Yang--Mills coupling constant. 
Quarks couple to dual potentials via a Dirac string connecting
the quark-antiquark pair along a line L, the ends of which 
are sources and sinks of color electric flux. The color field of the 
pair destroys the dual Meissner effect near L so that $\phi$
vanishes on L. At distances transverse to L greater than $1/M$
the monopole field returns to its bulk value $\phi_0$ so that 
 the color field is confined
to a tube of radius $a = 1/M$ surrounding the line L.
~\cite{Baker+Ball+Zachariasen:1991,Nora}  

As a result, 
for quark-antiquark separations $R$ greater than $a$, 
a linear potential develops that confines the quarks in hadrons.
At small separations, on the other hand, the color field generated
by the quarks expels the monopole condensate from the region
between them, and the potential becomes perturbative one gluon exchange.
The dual theory thus gives a potential that evolves smoothly
from the large $R$ confinement region to the short distance 
perturbative domain 
and provides an 
effective theory of long distance QCD, with short distances cut off
at the flux tube radius.
It does not describe QCD at shorter
distances where radiative corrections due to asmyptotic freedom
and a running coupling constant are important.

The dual theory must have classical vortex solutions which for short
distance reduce to perturbative QCD, and it must not contain massless
particles. For $SU(N)$ gauge theory this can be done
by coupling dual non-Abelian potentials $C_\mu$ to 3 multiplets
of scalar monopole fields $\phi$ in the adjoint representation,
so that in the confined phase the dual $SU(N)$ gauge symmetry is "spontaneously broken"
to $Z_N$. The gauge bosons $C_\mu$ then all acquire a mass, and 
there are
 $Z_N$ electric flux tube excitations~\cite{Baker+Ball+Zachariasen:1991,Nora} confining quark-antiquark pairs attached to their ends. Here we do not use a specific form for the action 
$S[C_\mu,\phi]$ 
of the dual theory.

 We want to use this theory to obtain the energy levels
of mesons having large angular momentum.  
Under such circumstances the distance between quarks is much
larger than the flux tube radius and we must include
the contribution of flux tube fluctuations to the interaction between
the quark and antiquark. The fluctuating vortex line L  
sweeps out a world sheet $\tilde x^\mu$  whose  boundary is
the loop $\Gamma$ formed from the world lines  of the moving 
quark-antiquark pair.  Their interaction is determined by 
the Wilson loop $W[\Gamma]$,
\begin{equation}
W[\Gamma] = \int \scrD C_\mu \scrD\phi \scrD\phi^* e^{iS[C_\mu, \phi]}
 \,.
\label{Wilson loop def}
\end{equation}
The path integral \eqnlessref{Wilson loop def} goes over all field
configurations for which the monopole field  $\phi(x)$
vanishes on some sheet $\tilde x^\mu$ bounded by the loop
$\Gamma$.

\section{The Effective String Theory}
 
The Wilson loop $W[\Gamma]$ describes the quantum fluctuations
of a field theory having classical vortex solutions. We want to
 express the functional integration over 
fields as a path integral over the vortex sheets $\tilde x^\mu$  
to obtain an effective string theory of these vortices. To do this
we carry out the functional integration \eqnlessref{Wilson loop def} 
in two stages:

\begin{enumerate}
\item We integrate over all field configurations
in which the vortex is located on a particular surface
$\tilde x^\mu$, where $\phi(\tilde x^\mu) = 0$. 
This integration determines the action $S_{{\rm eff}}[\tilde x^\mu]$
of the effective string theory.
\item We integrate over all
vortex sheets $\tilde x^\mu(\xi)$, $\xi=\xi^a$, $a=1\,,2$. 
This integration goes over the amplitudes $f^1(\xi)$ and $f^2(\xi)$
of the two transverse fluctuations of the world 
sheet $\tilde x^\mu(\xi)$ and 
gives $W[\Gamma]$ 
the form~\cite{Baker+Steinke2},
\begin{equation}
W[\Gamma] = \int \scrD f^1 \scrD f^2 \Delta_{FP} e^{iS_{{\rm eff}}[\tilde x^\mu]} \,,
\label{Wilson loop eff}
\end{equation}
where $\Delta_{FP}$ is a Faddeev-Popov determinant.
\end{enumerate}
The path integral \eqnlessref{Wilson loop def} over fields
has been transformed into the path integral 
\eqnlessref{Wilson loop eff}
over vortex fluctuations.
The factor $\Delta_{FP}$ in
\eqnlessref{Wilson loop eff}
arises from writing the original field theory path integral 
\eqnlessref{Wilson loop def} as a 
ratio of path integrals of two string theories~\cite{ACPZ}, and 
reflects the field theory 
origin of the effective string theory \eqnlessref{Wilson loop eff}.  
This ratio is anomaly~\cite{Polyakov:book} free.
The path integral \eqnlessref{Wilson loop eff} goes over 
fluctuations in the shape of the vortex sheet with wave lengths
greater than the flux tube radius.

The presence of the determinant $\Delta_{FP}$ makes the
path integral \eqnlessref{Wilson loop eff} 
invariant under reparameterizations $\tilde x^\mu(\xi)
\to \tilde x^\mu(\xi(\xi^\prime))$ of the vortex world sheet. 
The resulting  parameterization invariant measure in the path 
integral \eqnlessref{Wilson loop eff} is
independent of the explicit form of the dual field theory.
On the other hand, the action $S_{{\rm eff}}[\tilde x^\mu]$ 
of the effective string theory is 
not universal and depends upon the parameters of
the field theory.
However, for wave lengths of the string fluctuations greater
than the flux tube radius $a$
it can be expanded in powers of
the extrinsic curvature of the world sheet.
In mesons of angular momentum $J$ the expansion 
parameter is of order $1/J$.
For mesons having large angular momentum
the action $S_{{\rm eff}}$ can then be approximated by the 
first term in this expansion, namely
by the Nambu--Goto action,
\begin{equation}
S_{{\rm eff}}[\tilde x^\mu] \approx S_{{\rm NG}} \equiv -\sigma \int d^4x \sqrt{-g} \,,
\label{Nambu Goto Action}
\end{equation}
where  $\sqrt{-g}$ is the square root
of the determinant of the induced metric $g_{ab}$,
\begin{equation}
g_{ab} = \frac{\partial \tilde x^\mu}{\partial \xi^a}
\frac{\partial \tilde x_\mu}{\partial \xi^b} \,,
\end{equation}
and the string tension $\sigma $ is determined by the parameters
of the underlying dual field theory.
In the next section we use the 
results~\cite{Baker+Steinke2,ron}
of a semiclassical expansion 
of this effective string theory to calculate Regge trajectories
of mesons.

\section{Regge trajectories of Mesons}

\begin {figure}
    \begin{center}
	\null \hfill \epsfbox{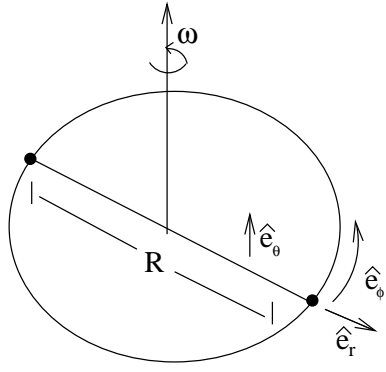} \hfill \null
    \end{center}
    \caption{The string coordinate system}
    \label{rot string fig}
\end {figure}

Consider a quark-antiquark pair rotating with uniform angular velocity $\omega$
(See Fig.~\ref{rot string fig}).
The quarks have masses $m_1$ and $m_2$,
move with velocities $v_1 = \omega R_1$ and $v_2 = \omega R_2$,
and are separated by a fixed distance $R = R_1 + R_2$. To describe
the string connecting this pair we use
parameters $\xi = (t, r)$ which are the time and the coordinate $r$,
which runs from $-R_1$ to $R_2$.

We evaluate $W[\Gamma]$ in the leading semiclassical approximation 
about a classical rotating straight string solution $\bar x^\mu(r,t)$.
The amplitudes $f^1(\xi)$ and $f^2(\xi)$ of the transverse fluctuations are
the spherical coordinates $\theta(r,t)$ and $\phi(r,t)$
of a point on the string. 
 The angle $\theta(r,t)$
is the fluctuation perpendicular to the plane of rotation, and
the angle $\phi(r,t)$ is the fluctuation lying in the plane
of the rotating string.  
The classical metric $\bar g_{ab} = g_{ab}[\bar x^\mu]$
and classical action $S_{{\rm NG}}[\bar x^\mu]$ are independent of
the time $t$, so that $W[\Gamma]$ has the form
\begin{equation}
W[\Gamma] = e^{iTL^{{\rm string}}(R_1, R_2, \omega)} \,,
\label{Wilson loop integrated}
\end{equation}
where $T$ is the elapsed time.

We calculate $L^{{\rm string}}$ in $D$
spacetime
dimensions. The fluctuation  $\theta(r,t)$ is replaced by 
$D-3$ fluctuations
perpendicular to the plane of rotation and there is still just 
1 fluctuation 
in the plane of rotation.
The effective Lagrangian for the quark--antiquark pair, obtained 
by adding
quark mass terms to $L^{{\rm string}}$,
is the sum of a classical part and a fluctuating part, \begin{equation} L_{{\rm eff}}(R_1, R_2, \omega) = L_{{\rm cl}} + L^{{\rm string}}_{{\rm fluc}}(R_1, R_2, \omega) \,, \end{equation} where \begin{equation} L_{{\rm cl}} = -\sum_{i=1}^2 m_i \sqrt{1-v_i^2} - \sigma \int_{-R_1}^{R_2} dr \sqrt{1 - r^2 \omega^2} \,,  \label{L_cl def formal} \end{equation} and the expression for $L^{{\rm string}}_{{\rm fluc}}$ is obtained from the semiclassical calculation of $W[\Gamma]$. It contains terms which are quadratically,
linearly, and logarithmically divergent in the cutoff $M$.
The quadratically divergent term is a renormalization of the
string tension, the linearly divergent term is a renormalization
of the quark mass, and the logarithmically divergent term is
a renormalization of a term in the boundary action called
the geodesic curvature \cite{Alvarez:1983}.
After absorbing these divergent terms into renormalizations,
we obtain \cite{Baker+Steinke2,ron}   
an expression 
for $L^{{\rm string}}_{{\rm fluc}}$
which remains finite in the massless quark limit,
\begin{equation}
L^{{\rm string}}_{{\rm fluc}}(\omega) \Big|_{m_1 = m_2 = 0}
= \frac{\omega (D-2)}{24} + \frac{\omega }{2} \,.
\label{L fluc ll}
\end{equation}
The first term in \eqnlessref{L fluc ll} is the negative of L\"uscher
potential with the length $R$ of the string replaced by its proper length 
$\pi/\omega$. It is the contribution of $D-2$ tranverse fluctuations
in the background of a flat metric. The $\frac{\omega}{2}$ term accounts
for the curvature of the classical background metric 
$\bar g_{ab}$ generated by the rotating string.
(For massless quarks, the ends of the
string move with the velocity of light and singularities
appear in $L^{{\rm string}}_{{\rm fluc}}$. 
The scalar quark mass term in $L_{{\rm eff}}$  
serves as an infrared regulator,
and in the limit of massless quarks gives no contribution to
$L^{{\rm string}}_{{\rm fluc}}$).

To take into account the fluctuations of the 
positions ${\bf \vec x}_1(t)$
and ${\bf \vec x}_2(t)$ of the quarks at the ends of the rotating 
string, we extend the functional
integral \eqnlessref{Wilson loop eff} to
include a path integral over ${\bf \vec x}_1(t)$ and
${\bf\vec x}_2(t)$. 
Using the methods of Dashen, Hasslacher and Neveu~\cite {DHN} to carry
out a semiclassical calculation of the path integral
gives the WKB quantization condition,
\begin{equation}
J = \frac{d L_{{\rm eff}}(\omega)}{d\omega} =
 l + \frac{1}{2} \kern 1 in l = 0, 1, 2,... \,.
\label{J quant}
\end{equation}
Furthermore, 
for massless quarks, there is no contribution to $L_{{\rm eff}}(\omega)$
arising from the fluctuations in the
motion of the ends of the string, 
so that $L_{{\rm fluc}}(\omega) = L^{{\rm string}}_{{\rm fluc}}(\omega)$.

Eqs' \eqnlessref{L_cl def formal} and \eqnlessref{L fluc ll} give
 $L_{{\rm fluc}}(\omega) / L_{{\rm cl}}(\omega) 
\sim \omega^2/\sigma \sim 1/J$,
so that for large $J$, $L_{{\rm
fluc}}(\omega)$ can be treated as a perturbation. The meson energy
is then
\begin{equation}
E(\omega) = E_{cl}(\omega) - L_{{\rm fluc}}(\omega)
= \frac{\pi\sigma}{\omega} - \frac{D-2}{24} \omega -\frac{\omega}{2} \,.
\label{E of omega ll}
\end{equation}

The energies $E_n(\omega)$ of the excited states of the rotating string
(light hybrid mesons) are
obtained by adding  to
\eqnlessref{E of omega ll} the energies $k\omega$ of the excited
vibrational modes.
\begin{equation}
E_n(\omega) = \frac{\pi\sigma}{\omega} - \frac{D-2}{24} \omega
-\frac{\omega}{2} + n \omega \,.
\label{hybrid energy ll}
\end{equation}
Since there are many combinations of string normal modes
which give the same $n$ (e.g., a doubly excited $k=1$ mode
and a singly excited $k=2$ mode each give $n=2$), the spectrum
is highly degenerate.

 The value of $\omega$ is given as a function of $J$ through the
classical relation $\omega = \sqrt{\pi\sigma/2J}$. Squaring
both sides of \eqnlessref{hybrid energy ll} and using
the WKB quantization condition \eqnlessref{J quant} yield
\begin{equation}
E_n^2  = 2\pi\sigma\left(l + n - \frac{D-2}{24} + O\left(
\frac{n^2}{l}\right)\right) \, .
\label{E^2 l}
\end{equation}
Setting $D=4$ in \eqnlessref{E^2 l}  and  solving for $l$ we obtain the
the  Regge trajectories,
\begin{equation}
l = \frac{E^2}{2\pi\sigma} + \frac{1}{12} - n
 \kern 1 in n =0, 1, 2, ... \,.
\label{hybrid l of E ll}
\end{equation}
The first semiclassical correction to the leading
Regge trajectory, $n=0$, adds the constant $1/12$ to the
classical Regge formula. 
For $D=26$, Eq. \eqnlessref{E^2 l}
yields the spectrum
\begin{equation}
E^2 = 2\pi\sigma\left(l+n - 1 
+O\left({n^2}/l \right)\right) \, .
\label{26 dim spectrum}
\end{equation}
The spectrum of energies \eqnlessref{26 dim spectrum}
, valid in the leading semiclassical approximation, 
coincides with
the spectrum of open strings in classical bosonic string theory. 

\section{Summary and Conclusions}

\begin{enumerate}
\item We have seen how the dual superconducting 
model of confinement leads to an effective string theory of 
long distance QCD. We have evaluated, in the semiclassical
large angular momentum 
 domain, where 
this theory is applicable, 
the contribution of string fluctuations to Regge trajectories
of mesons containing massless scalar quarks. The quark degrees
of freedom did not contribute to the meson energy. (With D. Gromes we are now 
investigating Regge trajectories of mesons formed by coupling 
spin 1/2 quarks to the string. We
find that, in the limit
that the quark masses go to zero, there are no finite energy states.
 This is another indication that
chiral symmetry must be broken in the confined phase of QCD.)
\item The derivation of the effective string theory
made no use of the details of the effective field theory
from which it was obtained. Furthermore, L. Yaffe has given arguments based on 
the work of 't Hooft~\cite{tHooft2} showing that the confined phase
of a non-Abelian gauge
theory is characterized by a dual order parameter, which
vanishes in regions of space where "dual supercondutivity" 
is destroyed. This generic description of the confined phase of QCD
leads to the effective string theory of long distance
QCD described here, which provides a concrete
picture of the QCD string.
\end{enumerate}



\end{document}